\documentclass[aps,preprint,nofootinbib,superscriptaddress,prc]{revtex4}
\usepackage{amssymb}
\usepackage{CJK}
\usepackage{indentfirst}
\usepackage{amsmath}

\usepackage{epsfig}
\usepackage[bookmarksnumbered,bookmarksopen,colorlinks,citecolor=blue,linkcolor=blue] {hyperref}
\begin{document}

\date{\today}

\title{Nuclear mass parabola and its applications}

\author{Junlong Tian}
 \email{tianjl@163.com}
 \affiliation{School of Physics and Electrical Engineering,
Anyang Normal University, Anyang 455000, People's Republic of
China }

\author{Di Yuan}
\affiliation{School of Physics and Electrical Engineering,
Anyang Normal University, Anyang 455000, People's Republic of
China }

\author{Yunyi Cui}
\affiliation{School of Physics and Electrical Engineering,
Anyang Normal University, Anyang 455000, People's Republic of
China }

\author{Yun Huang}
\affiliation{School of Physics and Electrical Engineering,
Anyang Normal University, Anyang 455000, People's Republic of
China }

\author{Ning Wang}\email{wangning@gxnu.edu.cn}
\affiliation{ Department of Physics, Guangxi Normal University,
Guilin 541004, People's Republic of China }
\affiliation{ State Key
Laboratory of Theoretical Physics, Institute of Theoretical Physics,
Chinese Academy of Sciences, Beijing 100190, People's Republic of
China}

\begin{abstract}
We propose a method to extract the properties of the isobaric mass parabola based on the total double $\beta$ decay energies of isobaric
nuclei. Two important parameters of the mass parabola, the location of the most $\beta$-stable nuclei $Z_{A}$ and the curvature parameter $b_{A}$, are obtained for 251 A values based on the total double $\beta$ decay energies of nuclei compiled in AME2016 database.
The advantage of this approach is that we can remove the pairing energy term $P_{A}$ caused by odd-even variation, and the mass excess $M(A,Z_{A})$ of the most stable nuclide for mass number $A$ in the performance process, which are used in the mass parabolic fitting method. The Coulomb energy coefficient $a_{c}=0.6910$ MeV is determined by the mass difference relation of mirror nuclei, and the symmetry energy coefficient is also studied by the relation $a_{\rm sym}(A)=0.25b_{A}Z_{A}$.
\end{abstract}


\maketitle

\section{\label{sec:level2}Introduction}
The nuclear mass (binding energy) is one of the most important topics in nuclear physics and astrophysics. In semi-empirical mass formula based on the liquid drop model, the binding energy $B(A,Z)$ of a nucleus is the summation of the volume, surface, Coulomb, symmetry and pairing terms, which can be expressed as a function of mass number $A$ and charge number $Z$ \cite{Weiz35,Bethe36},
\begin{eqnarray}
B(A,Z)=a_{\rm v}A-a_{\rm s}A^{2/3}-a_{\rm c}\frac{Z^{2}}{A^{1/3}}-a_{\rm sym}\frac{(N-Z)^{2}}{A}+a_{\rm p}A^{-1/2},
\end{eqnarray}
where a$_{v}$, a$_{s}$, a$_{c}$, a$_{\rm sym}$ and a$_{p}$ are the volume, surface, Coulomb, symmetry and pairing energy coefficients, respectively. The coefficients are obtained by fitting to experimentally measured binding energies of nuclei. On the other hand, the nuclear mass is usually indicated by the mass excess (nuclidic mass minus mass number in atomic mass units). The mass excess of the isobaric nuclei can be expressed as a parabola against the atomic number $Z$ with a high degree of approximation, which is called as a Bohr-Wheeler parabola \cite{Bohr39}. In Ref.\cite{Wing64}, the nuclidic mass equation is expressed as
\begin{eqnarray}
M(A,Z)= M(A,Z_{A})+\frac{1}{2}b_{A}(Z-Z_{A})^{2}+P_{A}-S(N,Z),
\end{eqnarray}
Where $M$ is the mass excess of the nucleus (A,Z) in MeV, $Z$ and $A$ are the atomic number and the mass number of the nucleus, respectively.
$M(A,Z_{A})$ is described as the minimum of the mass excess parabola, with $Z_{A}$ being a parameter and not the atomic number of an existing nuclei.
$b_{A}$ is the curvature parameter of the isobaric mass parabola,
$P_{A}$ is the pairing energy due to the odd-even variation, and $S(N,Z)$ is the shell correction term.
In the textbook \cite{Bauer11}, the relation between the nuclear mass excess and the binding energy is written as
\begin{eqnarray}
B(A,Z)=931.4943\times(0.008665A-0.00084Z)-M(A,Z),
\end{eqnarray}
where 0.008665 is the mass excess of neutron and 0.00084 is the hydrogen-neutron mass difference in atomic mass unit, and
one atomic mass unit is equivalent to 931.4943 MeV. Then whether the Coulomb energy coefficient and the symmetry energy coefficient in binding energy formula can be expressed by the parabolic parameters $Z_{A}$ and $b_{A}$?

Firstly, the two important parabolic parameters $Z_{A}$ and $b_{A}$ need to be determined. There are two ways to obtain these two parameters $Z_{A}$ and $b_{A}$, One is fitting to experimentally measured masses excess of isobaric nuclei with the Eq.(2). The other one is the fitting by measured beta-decay energies of  isobaric nuclei with the Eq.(4). The first mentioned of two methods need four parameters $Z_{A}$ ,$b_{A}$, $M(A,Z_{A})$ and $P_{A}$ to be determined by parabola fitting, the latter need three parameters $Z_{A}$ ,$b_{A}$ and $P_{A}$ to be determined by fitting to the $\beta$-decay energies.
Isobaric analysis of $\beta$-decay energies have been made by several authors \cite{Bohr39,Dewd63,LiXY81}, the early work being based on scanty experimental data. In Ref. \cite{Bohr39}, Bohr and Wheeler had used the least squares method to determine the parameters of the mass parabola, $Z_{A}$, $b_{A}$ and $P_{A}$ for 20 A values by parabola fitting to the isobaric mass excess. In Ref. \cite{Dewd63}, J. W. Dewdney had analyzed the isobaric $\beta$ decay energies and used the least squares method to determine the parameters of the mass parabola for 157 A values. The experimental data are taken from the Refs. \cite{Ever60,Koning62}. Later, X. Y. Li and co-workers \cite{LiXY81} updated the three parameters of isobaric mass parabola for 234 A values ($10\leq A\leq 253$) in the same manner as Dewdney but adopted the different mass table AME1977 \cite{AME77} including about 1000 nuclides. In recent decades, with the development of experimental instruments and the progress of science technology, a large number of unstable nuclei can be produced, and their masses can be measured with high precision. The experimental information or recommended values for nuclear and decay properties of 3437 nuclides are compiled in the mass table of the AME2016 \cite{AME16}. The number and precision of nuclear mass in AME2016 have highly increased in comparison with the results compiled in the AME1977 database \cite{AME77}. The available data are now so much more extensive than ever before that it is possible to analyse them in a statistical way.

Double $\beta$ decay is also a popular topic, which is a rare transition between two nuclei with the same mass number that changes the nuclear charge number by two units, that is to say, $(A, Z) \rightarrow (A, Z\pm2)$ has been observed in many nuclei \cite{Rode11,Beri12}. The energy release in double $\beta$ decay shows even greater regularity than that of single $\beta$-decay as shown in Ref. \cite{Way54}, but detailed studies of the properties of the mass parabola have not been performed by using the total double $\beta$-decay energies.
In this paper, we propose a very simple empirical formula to obtain the parameters $Z_{A}$ and $b_{A}$. And then the relation of Coulomb energy coefficient and the symmetry energy coefficient with $Z_{A}$ and $b_{A}$ will be presented. So we use more than 2400 total double $\beta$ decay energies, which are complied in the AME2016 database, to analyze the properties of the Bohr-Wheeler isobaric mass parabola by all modes of double $\beta$ decay in the theoretical method. The advantage of this approach is that there are only two parameters including in the expression of the double $\beta$-decay energies, the pairing energy term $P_{A}$ and the mass excess $M(A,Z_{A})$ are removed by the mass difference between $(A, Z)$ and $(A, Z\pm2)$. Furthermore, the over-all simplicity of the double $\beta$ decay energy pattern may point the way to a convenient empirical mass formula.

\section{\label{sec:level1}Theoretical framework}

The $Q$ value is defined as the total energy released in a given nuclear reaction. The $Q$ value of $\beta$ decay is calculated by the mass difference between the two nuclei $(A, Z)$ and $(A, Z+1)$,
while the expressions for $\beta$ decay energies may be derived
from empirical mass equation Eq. (2) when shell correction be neglected. Following Eq.(2), the $\beta$ decay energies can be obtained which is directly proportional to $Z-Z_{A}$. For odd $A$ isobars, there is only one isobaric mass parabola due to the pairing energy term equal to zero. And the $\beta$ decay energies plotted versus $Z$ lie on a single straight line. But for even $A$ isobars, because of the extra stability associated with pairs of like nucleons, two isobaric mass parabolas with the same shape are produced but displaced one below the other. On the lower parabola lie the even-even nuclides, on the upper parabola lie the odd-odd nuclides. So the $\beta$ decay energies plotted versus $Z$ lie not on a single straight line but on a pair of parallel straight lines. For negative $\beta$ decay and positive $\beta$ decay (or electron capture), one can combine them into one universal description,
\begin{eqnarray}
Q_{\beta}=M(A,Z)-M(A,Z+1)=-b_{A}(Z-Z_{A}+\frac{1}{2})+\Delta E.
\end{eqnarray}
The reaction will proceed only when the Q value is positive. Where ``$Q_{\beta}>0$" for $\beta^{-}$ decay occurs when the mass of atom $M(A, Z)$
is greater than the mass of atom $M(A, Z+1)$;  ``$Q_{\beta}<0$" for $\beta^{+}$ decay occurs when the mass of atom $M(A, Z+1)$
is greater than that of $M(A, Z)$.
The $\Delta E=2P_{A}$ for even-even nuclides, the $\Delta E=-2P_{A}$ is for odd-odd
nuclides, and for odd-A nuclides (i.e. even-odd and odd-even)
$\Delta E=0$.

In this paper our aim is to propose a very simple empirical formula that only depends on the basic mass parabolic parameters $Z_{A}$ and $b_{A}$. So we use the total double $\beta$ decay energies to analyze the properties of the mass parabola rather than total single $\beta$ decay energies. The parameters $M (A,Z_{A})$ and $\Delta E$ are removed by the mass difference between $(A, Z)$ and $(A, Z + 2)$.
We can obtain the universal expression to describing total double $\beta$ decay energies in the same manner above mentioned,
\begin{eqnarray}
Q_{2\beta}=M(A,Z)-M(A,Z+2)=-2b_{A}(Z-Z_{A}+1),
\end{eqnarray}
where ``$Q_{2\beta}>0$" for $\beta^{-}\beta^{-}$, and ``$Q_{2\beta}<0$" for $\beta^{+}\beta^{+}$. The nuclei $(Z,A)$ will be called the decay nuclei no matter whether the decay actually proceeds from $(A, Z)$ to $(A, Z + 2)$ or vice versa. The total double $\beta$ decay energies lie on the same lines for odd-odd and even-even nuclide for even-A values, and for adjacent odd-A and even-A values lie approximately on the same lines.

\section{\label{sec:level1}The results and discussions}

\subsection{\label{subsec:leve1}The procedure of the calculated values $Z_{A}$ and $b_{A}$}

We use the three equations Eqs.(2),(4) and (5) to analyse the corresponding experimental data, i.e. mass excess, total $\beta$ decay energies and total double $\beta$ decay energies, both parameters $Z_{A}$ and $b_{A}$ of the isobaric mass parabola are obtained by using the least-squares fitting procedure. We obtain the optimal values $Z_{A}$ and $b_{A}$ are list in three attached files (see the Supplemental Material table 1-3.txt \cite{Supp3}). The calculated results are compared, and they are almost identical for three different methods to fit the corresponding experimental data.

But the first equation Eq.(2) at least need three parameters $Z_{A}$, $b_{A}$ and $M (A,Z_{A})$ to fit experimental mass excess for odd-A nuclei. Here, $P_{A}$ =0 is taken, in fact the values of $P_{A} <0.3$ MeV for odd-A nuclei presented in Ref. \cite{Dewd63}.
For even-$A$ nuclei, an isobaric slice through the mass surface tends to yield two parabolas of the same shape but displaced one below the other. On the lower parabola lie the even nuclides, on the upper parabola lie the odd nuclides. So it is necessary to unify the parameters $Z_{A}$ and $b_{A}$ for even-even nuclei and odd-odd nuclei, $b_{A}=0.5(b^{o-o}_{A}+b^{e-e}_{A})$ and $Z_{A}=0.5(Z_{A}^{o-o}+Z_{A}^{e-e})$ , where $b^{e-e}_{A}$ and $Z_{A}^{e-e}$ for even-even nuclei, $b^{o-o}_{A}$ and $Z_{A}^{o-o}$ for odd-odd nuclei, respectively. Finally, the value of $\Delta E$ is obtained by the difference of $\Delta E=M^{o-o}(A,Z_{A})-M^{e-e}(A,Z_{A})$. For Eq. (4), it is similar procedure to that of Eq. (2), the value of $\Delta E$ equals the half of the difference between two intercepts of a pair of parallel lines.

\begin{figure}
\includegraphics[angle=-0,width= 0.95\textwidth]{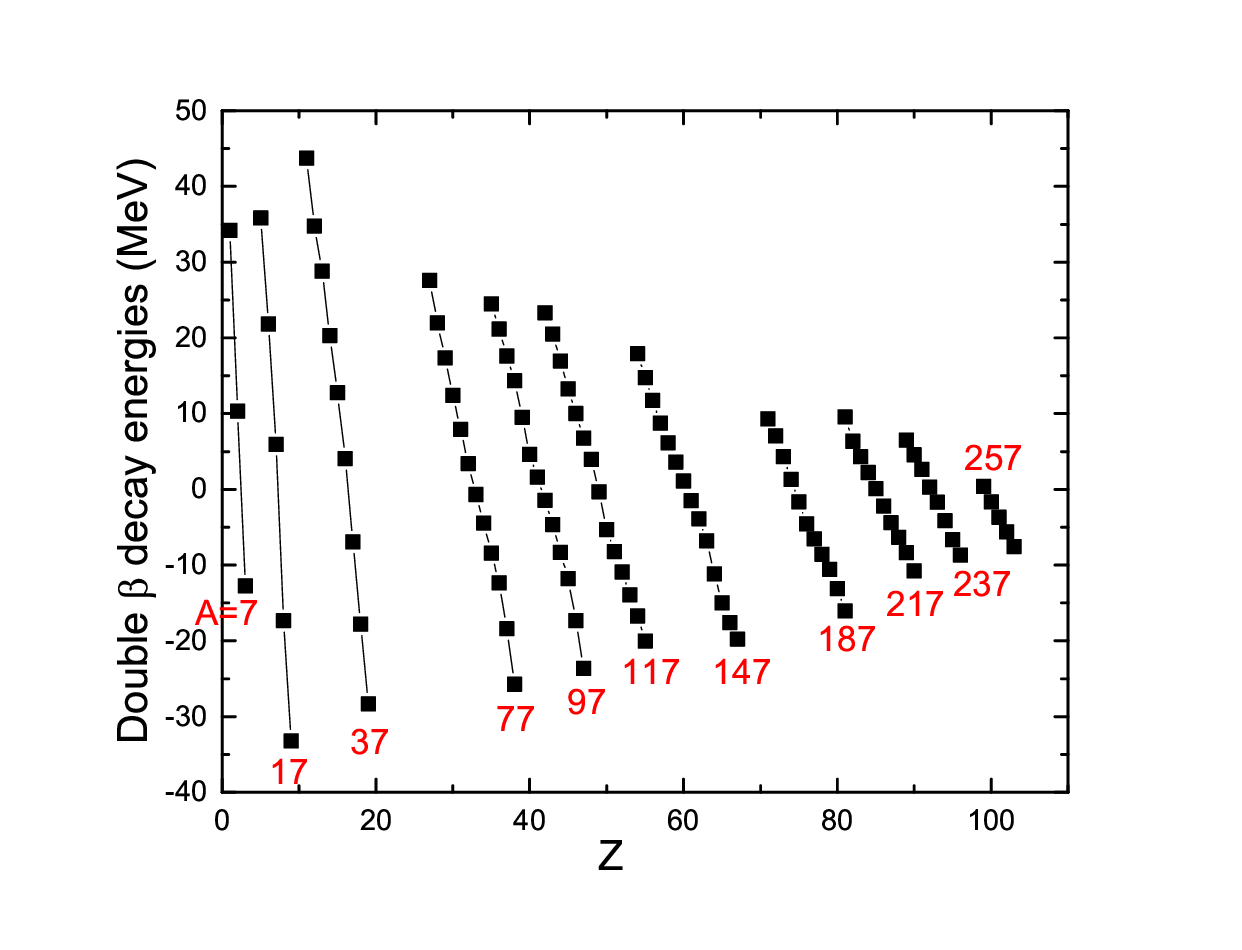}
 \caption{(Color online) Total double $\beta$ decay energies $Q_{2\beta}$ as a function of the charge number of $Z$ for 11 odd-$A$ values ($7\leq A\leq257$). The experimental data are taken from AME2016 \cite{AME16}.}
\end{figure}

For the construction of the $Q_{2\beta}$ formula we start
by plotting the total double $\beta$ decay energies $Q_{2\beta}$ versus the charge number Z for 11 odd-$A$ values $7\leq A\leq257$ in Fig.1. The experimental data of $Q_{2\beta}$ are taken from AME2016 \cite{AME16}. One can see clearly straight line relation in this plot, and the slopes $2b_{A}$ of these lines show the general trend of $Q_{2\beta}$ to decrease slightly with increasing mass number $A$. The similar linear relation is also shown for even-$A$ values in Fig. 2.

\begin{figure}
\includegraphics[angle=-0,width= 0.99\textwidth]{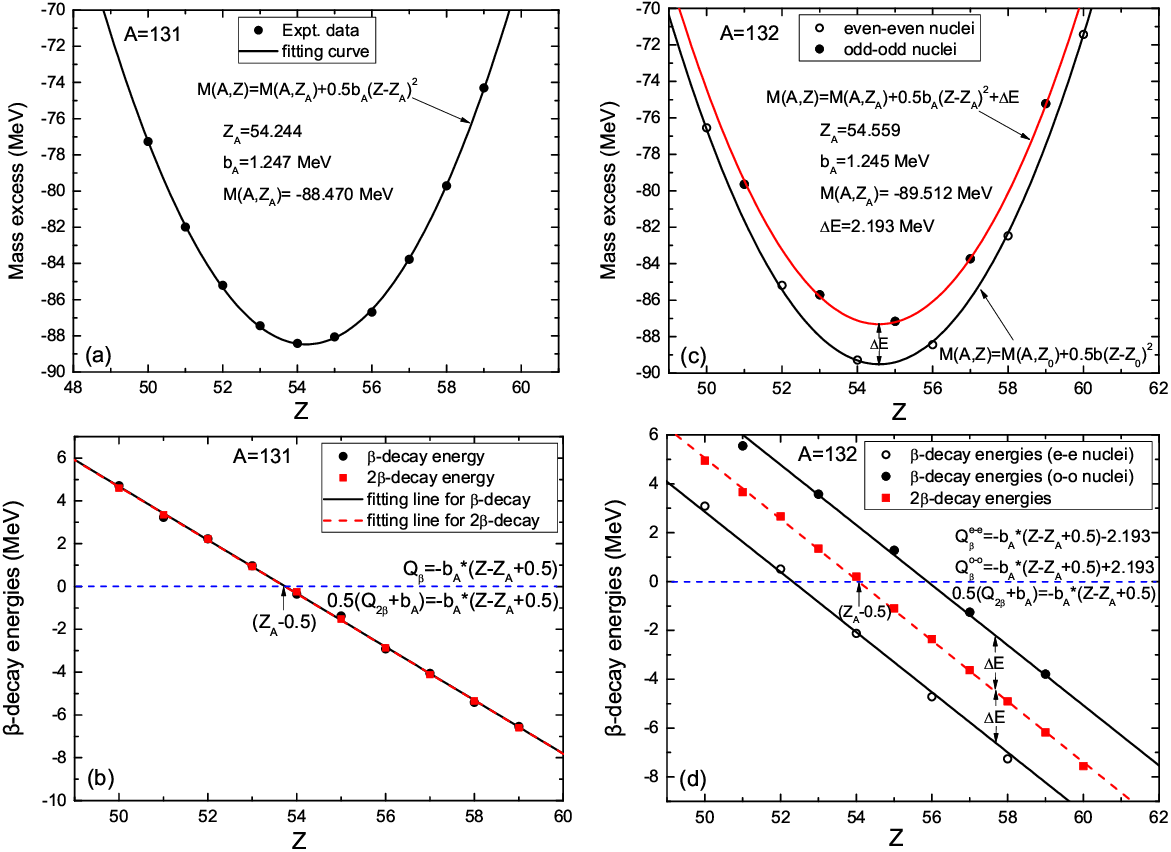}
 \caption{(Color online) Two examples of experimental data and their treatment. The mass excesses (a), the $\beta$ decay energies (b) and the curves fitted to them for $A=131$, (c) and (d) are same as (a) and (b) for $A=132$, as a function of the nuclear charge number $Z$. The solid curves in Fig. 2 (a) and (c) denote the results of the fitting experimental mass excess by Eq. (2). The solid lines in Fig. 2 (b) and (d) denote the results of the fitting experimental data $Q_{\beta}$ by Eq. (4), and the dashed lines in Fig. 2 (b) and (d) denote the results of the fitting experimental data $Q_{2\beta}$ by Eq. (5).}
\end{figure}

Fig. 2 shows two examples of the treatment of experimental data for $A=131$ and $A=132$ as a function of the nuclear charge number $Z$. The Fig.2 (a) is a plot of the experimental isobaric mass parabola at $A$ = 131. The mass parabola imply the $\beta$ or $2\beta$ decay energies directly below (b). The intercept $Z_{A}$ on the $\beta$ decay energy graph corresponds to the minimum of the mass parabola. The slope $b_{A}$ of the $\beta$ energy graph is a measure of the steepness of the mass parabola. The red dashed lines in Fig.2 (b) and (d) denote the results of the fitting experimental data $Q_{2\beta}$ by Eq. (5). The Fig.2 (c) and (d) are similar plots but for $A$ = 132. Fig. 2 (c) shown mass excesses lie alternately on a pair of parabolas of identical shape. The parabola on which the even masses lie falls a distance $\Delta E$ below the odd parabola. This implies that $\beta$ decay points lie on a pair of parallel straight lines, while the double $\beta$ decay points lie on a single straight line for both odd-A and even-A values.

\begin{figure}
\includegraphics[angle=-0,width= 0.75\textwidth]{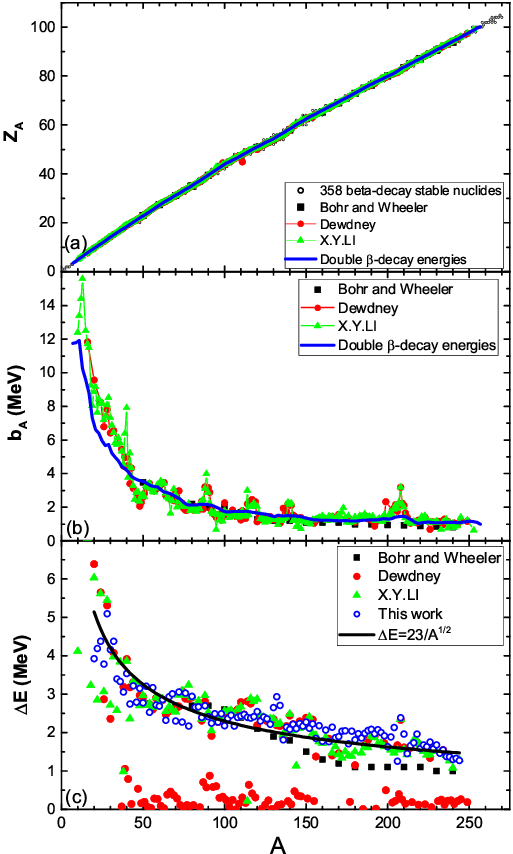}
 \caption{(Color online) The mass parabola parameters $Z_{A}$ (a), $b_{A}$ (b) and $\Delta E$ (c) as a function of nuclear mass number $A$. The calculated values (the blue solid curves) are compared with early results of Bohr and Wheeler \cite{Bohr39} (the solid squares), Dewdney \cite{Dewd63} (solid circles) and X.Y.Li \cite{LiXY81} (solid triangles). The black solid curve in (c) is the pairing energy term $\Delta E=2P_{A}$ which is obtained by fitting the $\beta$ decay energies.}
\end{figure}

Fig. 3 presents the mass parabolic parameters $Z_{A}$ (a), $b_{A}$ (b) and $\Delta E$ (c) as a function of nuclear mass number $A$. The values of $Z_{A}$ and $b_{A}$ we obtained through the aforementioned procedure are shown in Fig. 3 (a) and (b) with the blue solid curves by using Eq. (5). The $\Delta E$ are plotted with blue open circles in Fig. 3 (c) calculated by Eq.(4). The solid curve $\Delta E=2P_{A}$ is the result to fit the $\beta$ decay energies, and the pairing energy term $P_{A}=11.50\delta/A^{1/2}$ is obtained, where $\delta$ equals 0 for odd-A, +1 for even-even and -1 for odd-odd nuclides. The calculated values are compared with early results of Bohr and Wheeler \cite{Bohr39} (the solid squares), Dewdney \cite{Dewd63} (solid circles) and X.Y.Li \cite{LiXY81} (solid triangles). There are currently known to be 358 beta-decay stable nuclides \cite{Stable358} plotted in Fig. 3 (a) with open circles. We found a very good agreement between our results and those obtained previously in Ref. \cite{Dewd63,LiXY81} with exception of few A values (A=111 for $Z_{A}$ parameter and A=144 for $b_{A}$ parameter). We think that for these cases our results are more robust, because the marked discontinuities occur only at the two A values shown in Fig. 3. Our results are $Z_{A}$=47.7566, $b_{A}$=1.7955 for A=111, and $Z_{A}$=59.9768, $b_{A}$=1.5102 for A=144.

\subsection{\label{sec:level1} Determination of the Coulomb energy coefficient $a_{c}$ by the relation of $Z_{A}$ and $b_{A}$}
The Coulomb energy coefficient $a_{c}$ can be determined by the mass relation of mirror nuclei. On the one hand, due to the charge-independence of nuclear force, the binding energies of mirror pairs differ only in their Coulomb energies. The Coulomb energy difference between a pair of mirror nuclei is proportional to $Y=N-Z$ \cite{Tian2013,Ormand97},
\begin{eqnarray}
-\Delta B = E_{C}[A,\frac{1}{2}(A+Y)]-E_{C}[A,\frac{1}{2}(A-Y)] = b_{c}Y,
\end{eqnarray}
in which $b_{c}$ is the proportionality coefficient which indeed depends on $A$.

On the other hand, the mirror mass relations can be obtained by Eq.(2).
Firstly, one need to express charge number $Z$ with mass number $A$, that is to say, $Z=\frac{1}{2}(A\pm Y)$ for a pair of mirror nuclei.
Then the difference of mass excess for a pair mirror nuclei for both odd and even values of $A$ with
different $Y=1, 2, 3,...$ reads
\begin{eqnarray}
\Delta M=M[A,\frac{1}{2}(A+Y)]-M[A,\frac{1}{2}(A-Y)]=\frac{1}{2}b_{A}(A-2Z_{A})Y.
\end{eqnarray}
Combining the three equations Eqs. (3),(6) and (7), and taking the hydrogen-neutron mass difference $\Delta_{(^{1}H-n)}=-0.7825$ MeV, we can obtain
\begin{eqnarray}
\frac{\Delta M}{Y}=b_{c}=\frac{1}{2}b_{A}(A-2Z_{A})-\Delta_{(^{1}H-n)}.
\end{eqnarray}

\begin{figure}
\includegraphics[angle=-0,width= 0.95\textwidth]{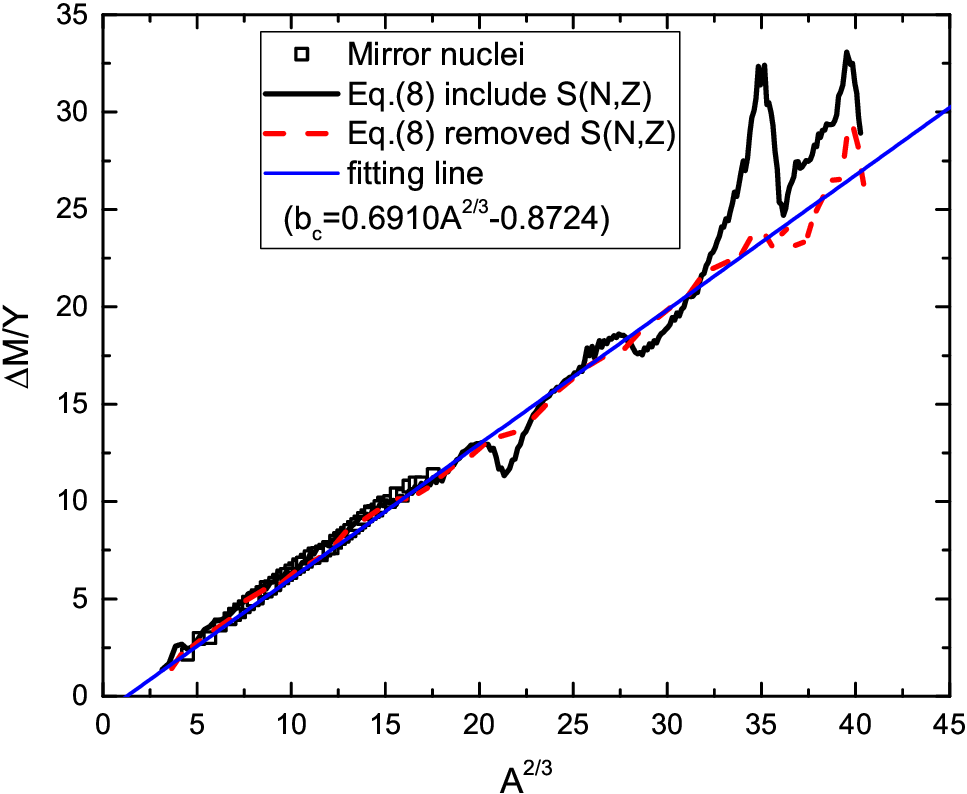}
 \caption{(Color online) Scaled mass difference (solid squares) of 95 pairs of mirror nuclei in the region 11$\leq A\leq$75 as a function of $A^{2/3}$. The solid curve and the dashed curve represent the left hand of Eq.(8) including the shell correction energies and removed the shell correction energies, respectively. The shell correction energies are taken from Ref. \cite{KTUY05}. The thin solid line is the fitting straight line.}
\end{figure}
This result is consistent with that in Refs. \cite{Durn69,Tian2014}.
Our results are presented in Fig. 4, the mass differences of the 95
pairs of mirror nuclei, scaled by the charge difference Y,
are plotted versus $A^{2/3}$ and are seen to lie on a straight
line. The value of Y ranges from 1 (32 cases) to 6 (1
case). The solid and dashed curves denote the results with and without shell correction energies, respectively.
The shell correction energies take from Ref. \cite{KTUY05}.
The thick solid curve shows some oscillations and fluctuations due to the shell effect. When the shell corrections are taken into account, the fluctuations in the extracted $b_{c}$ are reduced effectively.  The thin solid line is the fitting straight line. We obtain $b_{c}=0.691A^{2/3}-0.8724$ MeV by fitting left hand of Eq.(8) to 251 $A$ values with an
rms deviation of 0.384 MeV, the error associated with this method to determine $a_{c}$ and $c$ smaller than what done in Ref. \cite{Durn69}. The intercept 0.8724 is the contributions of the Coulomb exchange term and the nuclear surface diffuseness correction term. If adopt the Coulomb energy expression $E_{c}=\frac{a_{c}Z^{2}}{A^{1/3}}(1-cZ^{-2/3})$, in which the contributions of the Coulomb exchange term is taken into account, we obtain $a_{c}=0.691$ MeV and $c=1.1914$ MeV.
It is shown that the mirror-nuclide method can be extended to include all values of A in the Eq. (8), though mirror nuclides are observed for $A \leq 75$. The availability of the Coulomb energy coefficient for the
complete range of $A$ values should be helpful in a study of the variation of the Coulomb
energy with mass number.

\subsection{\label{sec:level1} Determination of the symmetry energy coefficient $a_{\rm sym}$ by the relation of $Z_{A}$ and $b_{A}$}
The symmetry energy coefficient plays a dramatic role in the nuclear physics and astrophysics. In this work, the symmetry energy coefficient $a_{\rm sym}$ is deduced and expressed by the two parabolic paremeters $Z_{A}$ and $b_{A}$. Firstly, we insert Eq.(1) and Eq.(3) into the expression Eq.(5) and take the double $\beta$ decay Q value of the mother nuclide (A,Z-1), one can obtain
\begin{eqnarray}
\nonumber Q_{2\beta}(A,Z-1) &&=M(A,Z-1)-M(A,Z+1)\\
\nonumber  &&=B(A,Z+1)-B(A,Z-1)+2\Delta_{(^{1}H-n)}\\
 &&=-(\frac{16a_{\rm sym}}{A}+\frac{4a_{c}}{A^{1/3}})Z+8a_{\rm sym}+1.5649.
\end{eqnarray}
While the double $\beta$ decay Q value of (A,Z-1) in Eq.(5) is written as
\begin{eqnarray}
Q_{2\beta}(A,Z-1)=-2b_{A}(Z-Z_{A}),
\end{eqnarray}
Then the symmetry energy coefficient $a_{\rm sym}$ is obtained by solving the combination of Eqs.(9) and (10), that is
\begin{eqnarray}
a_{\rm sym} = \frac{(2b_{A}Z_{A}-1.5649)}{8}\simeq\frac{b_{A}Z_{A}}{4}.
\end{eqnarray}

\begin{figure}
\includegraphics[angle=-0,width= 0.95\textwidth]{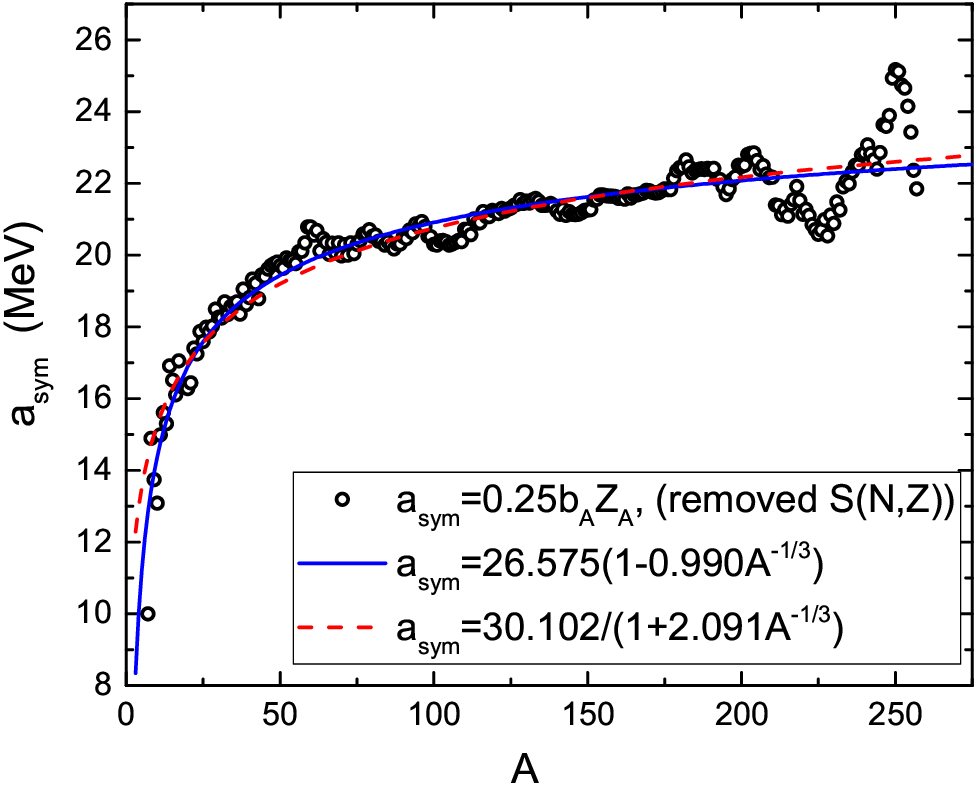}
 \caption{(Color online) The symmetry energy coefficient of finite nuclei
$a_{\rm sym}(A)$ as a function of mass number A from Eq. (11) (open circles), but the shell
corrections S(N,Z) from Ref. \cite{KTUY05} are removed from the mass excess of nuclei.
The blue solid curve and the red dashed curve denote the results of two analytic expressions in
which the coefficients are determined by fitting the open circles.}
\end{figure}
There are two kinds form for description of the mass-dependent symmetry coefficient. One is written by Danielewicz and Lee as
$a_{\rm sym}(A)=S_{0}/(1+\kappa A^{-1/3})$ \cite{Deniel09}, where $S_{0}$ is the volume symmetry energy coefficient of the
nuclei and $\kappa$ is the ratio of the surface symmetry coefficient to the volume symmetry coefficient. The other form is frequently used and
written as $a_{\rm sym}(A)=S_{0}(1-\kappa A^{-1/3})$ \cite{Moller95,Stoi07,Satu06,Sama07,Kolo10,Jiang12}.
Fig. 5 shows the symmetry energy coefficient $a_{\rm sym}(A)$ as a function of A from Eq. (11). The open circles denote the results without the shell corrections. But it also shows some odd-even staggerings and fluctuations, they can be caused by nuclear residual pairing interaction and the nuclear surface diffuseness correction effect. The blue solid curve and the red dashed curve denote the results of two analytic expressions in which the coefficients are determined by fitting the open circles. By performing a two-parameter fitting to the $a_{\rm sym}(A)$ obtained previously by the relation Eq.(11), we have obtained the values of $S_{0}$  and $\kappa$. With 95\% confidence intervals, we obtain the values of $S_{0}=26.575\pm0.271$ and $\kappa=0.99\pm0.035$, and the corresponding rms deviation is 389 keV, if assuming mass dependence of symmetry energy coefficient $a_{\rm sym}(A)=S_{0}(1-\kappa A^{-1/3})$. By adopting $a_{\rm sym}(A)=S_{0}/(1+\kappa A^{-1/3})$, one obtains $S_{0}=30.102\pm0.741$ MeV and $\kappa=2.091\pm0.166$, respectively, with an rms deviation of 398 keV. The result is shown in Fig.5 by the red dashed curve. The obtained values of $S_{0}$ and $\kappa$ are in agreement with the range of $S_{0}=31.1\pm1.7$ MeV and $\kappa=2.31\pm038$ given by Min Liu \cite{Liu10}.

\section{\label{sec:level1}Summary}

In summary, we have proposed a method to determine the two important parameters $Z_{A}$ and $b_{A}$ in the well-known Bohr-Wheeler mass parabola. The linear relation between the total double $\beta$ decay energies and charge number $Z$ is deduced based on the isobaric mass parabola, which only include two parameters of $Z_{A}$ and $b_{A}$. The values of these two parameters have been refitted for 251 A values based the latest experimental total double $\beta$ decay energies of nuclei compiled in AME2016 database. The advantage of this approach is that we can remove the pairing energy term $P_{A}$ caused by odd-even variation, and the mass excess $M(A,Z_{A})$ of the most stable nuclide in the performance process. The Coulomb energy coefficient $a_{c}=0.691$ MeV has been determined by the mass difference relation of mirror nuclei $0.5b_{A}(A-2Z_{A})=\Delta_{(^{1}H-n)}+a_{c}(A^{2/3}-1.0583c)$, and $c=1.1914$ MeV. The symmetry energy coefficients of the mass dependence under two different form, $a_{\rm sym}(A)$, $a_{\rm sym}(A)=30.102/(1+2.091A^{-1/3})$ and $a_{\rm sym}(A)=26.575(1-0.99A^{-1/3})$, have been determined by fitting the relation $a_{\rm sym}(A)=0.25b_{A}Z_{A}$. The obtained values of $S_{0}$ and $\kappa$ are in agreement with the results in references. It implies that the method is reliable to determine Coulomb energy coefficient and the symmetry energy coefficient. The further work is in progress which include the comparison of the results between the experimental data and several theoretical mass tables.

\begin{center}
\textbf{ACKNOWLEDGEMENTS}
\end{center}
This work was supported by National Natural Science Foundation of China, Nos. 11961131010,
11422548 and U1867212, and the Foundation of Guangxi innovative team (Nos. 2017GXNSFGA198001, 2016GXNSFFA380001), and innovation fund of undergraduate at Anyang Normal
University (ASCX/2019-Z055).


\begin{thebibliography}{99}
\bibitem{Weiz35} C. F. Von Weizs$\ddot{a}$ker, Z. Phys. 96, 431, (1935) .
\bibitem{Bethe36} H. A. Bethe and R. F. Bacher, Rev. Mod. Phys. 8, 82 (1936).
\bibitem{Bohr39} N. Bohr and J. A. Wheeler, Phys. Rev. \textbf{56},426 (1939).
\bibitem{Wing64} J. Wing and P. Fong, Phys. Rev. \textbf{136},B932 (1964).
\bibitem{Bauer11} W. Bauer and G. D. Westtfall, University Physics with modern Physics, McGraw Hill, 1330 (2011).
\bibitem{Dewd63} J. W. Dewdney, Nucl. Phys. 43, 303 (1963).
\bibitem{LiXY81} Xian-Yin Li, Shi-Huai Yao and Fu-Xin Xu,  Chin. Phys. C 5, 611 (1981).
\bibitem{Ever60} F. Everling, L. A. K$\ddot{o}$nig, J. H. E. Mattauch and A. H. Wapstra, Nuclear Physics 18, 529 (1960).
\bibitem{Koning62} L. A. K$\ddot{o}$nig, J. H. E. Mattauch and A. H. Wapstra, Nuclear Physics 31 (1962) 18
\bibitem{AME77} A. H. Wapstra and K. Bos, Atomic data and nuclear data tables,19, 175 (1977).
\bibitem{AME16} G. Audi, F. G. Kondev, M. Wang, W.J. Huang and S. Naimi, Chin. Phys. C 41, 030001 (2017).
\bibitem{Rode11} W. Rodejohann, Int. J. Mod. Phys. E 20, 1833 (2011).
\bibitem{Beri12} J. Beringer, \emph{et al}, (Particle Data Group), Phys. Rev. D 86, 010001 (2012).
\bibitem{Way54} K. Way and M. Wood, Phys. Rev. 94, 119 (1954).
\bibitem{Supp3} See Supplemental Material for obtained parameters of mass parabola by using Eqs.(2),(4),(5) in this paper.
\bibitem{Stable358} Interactive Chart of Nuclides (Brookhaven National Laboratory), http://www.nndc.bnl.gov
/chart/; or https://en.wikipedia.org/wiki/Beta-decay stable isobars.

\bibitem{Tian2013} J. L. Tian, N. Wang, C. Li, and J. J. Li, Phys. Rev. C \textbf{87}, 014313 (2013).

\bibitem{Ormand97} W. E. Ormand, Phys. Rev. C \textbf{55}, 2407 (1997).
\bibitem{Durn69} A. M. Durnford , J. Phys. A \textbf{2}, 59 (1969).
\bibitem{Tian2014} J. L. Tian, H. T. Cui, K.K. Zheng, and N. Wang, Phys. Rev. C 90, 024313 (2014).
\bibitem{KTUY05} H. Koura, T. Tachibana, M. Uno, and M. Yamada, Prog. Theor. Phys. 113,305 (2005).
\bibitem{Deniel09} P. Danielewicz and J. Lee, Nucl. Phys. A \textbf{818}, 36 (2009)
\bibitem{Moller95}P. M$\ddot{o}$ller, J.R. Nix, W.D. Myers, and W.J. Swiatecki, At. Data Nucl. Data Tables  \textbf{59}, 185 (1995).
\bibitem{Stoi07} M. Stoitsov, R. B. Cakirli, R. F. Casten, W. Nazarewicz, and W. Satula, Phys. Rev. Lett. 98, 132502 (2007).
\bibitem{Satu06} W. Satula, R. A. Wyss, and M. Rafalski, Phys. Rev. C 74, 011301(R) (2006).
\bibitem{Sama07} S. K. Samaddar, J. N. De, X. Vinas, and M. Centelles, Phys. Rev. C 76, 041602(R) (2007).
\bibitem{Kolo10} V. M. Kolomietz and A. I. Sanzhur, Phys. Rev. C 81, 024324 (2010).
\bibitem{Jiang12} H. Jiang, G. J. Fu, Y. M. Zhao, and A. Arima, Phys. Rev. C 85, 024301 (2012).
\bibitem{Liu10} M. Liu, N. Wang, Z. X. Li, and F. S. Zhang, Phys. Rev. C 82, 064306 (2010).


\end{thebibliography}
\end{document}